\documentclass[11pt]{article}
\newcommand{\be}{\begin{equation}}
\newcommand{\ee}{\end{equation}}
\newcommand{\beq}{\begin{eqnarray}}
\newcommand{\eeq}{\end{eqnarray}}

\def\theequation{\arabic{section}.\arabic{equation}}
\usepackage{graphicx}

\begin{document}
\begin{center}
{\bf\LARGE QED vacuum loops and Inflation }\\
[7mm]
\vspace{1cm}  
H. M. FRIED
\\
{\em Department of Physics \\
Brown University \\
Providence R.I. 02912 USA}\\
fried@het.brown.edu\\
[5mm]
Y. GABELLINI
\\
{\em Institut Non Lin\'eaire de Nice\\
UMR 7335 CNRS\\
 1361 Route des Lucioles\\
06560 Valbonne France}\\
yves.gabellini@unice.fr\\
[5mm]

\vspace{4cm}
Abstract
\end{center}
A QED--based model of a new version of Vacuum Energy has recently been suggested, which leads to a simple, finite, one parameter representation of Dark Energy. An elementary, obvious, but perhaps radical generalization is then able to describe both Dark Energy and Inflation in the same framework of Vacuum Energy. One further, obvious generalization then leads to a relation between Inflation and the Big Bang, to the automatic inclusion of Dark Matter, and to a possible understanding of the birth ( and death ) of a Universe.
\newpage
{\section{Introduction}}
\setcounter{equation}{0}

This paper defines an extension of a previous QED--based model of Dark Energy \cite{FG1}, which described a ``bootstrap'' solution for a vacuum energy following from the existence of fluctuating lepton and quark pairs in the quantum vacuum. That vacuum energy appeared with a very high frequency, on the order of the Planck mass $M_P$. Gauge and Lorentz invariance were easily incorporated; and even a cursory glance at the form of that vacuum energy density, figure 1, below, suggests application to both Dark Energy and Inflation.

However, there is one crucial difference between the QED mechanism which can generate a present day vacuum energy, associated with Dark Energy and that  which is suitable for Inflation, describing how our Universe evolved from a speck of infinitesimally small, space--time dimensions \cite{PS}. That distinction appears because present day lepton and quark pair fluctuations are described in terms of renormalized charge. But in order for charge renormalization to exist, one must be able to view the bare charge at distances larger than the Compton wavelength of the particle carrying the charge; and this is impossible in the context of the beginning of Inflation, where distances are on the order of the inverse of the Planck mass, $10^{-31}\,$cm. In order for this QED vacuum energy model to be applicable here, one must resort to working within a formalism that contains only the unrenormalized charge, $e_0$. Is this possible ?

In fact, it is quite possible, following the functional analysis of QED charge renormalization presented in \cite{FG2}, where the summation and inclusion of the contributions of an infinite number of loop functionals $L[A]$ suggested that charge renormalization is indeed finite; and to within the qualitative approximations of that estimate, indicate that the fine structure constant calculated with $e_0$, rather than the renormalized $e_R$ is given by $\alpha_0 = \pi/2$. This will be the value used when writing the unrenormalized ( but appropriately cut--off ) loop fluctuation $\Pi(k^2)$.

In order to formulate the QED--based model of ref.\cite{FG1} for Inflation, we are therefore committed to using $\alpha_0$, rather than $\alpha = 1/137$, in the unrenormalized contributions of the vacuum loops; and here a new difficulty arises, for the term corresponding to the unrenormalized $\Pi(k^2)$ is no longer real, but contains an imaginary part which increases as each lepton loop is added to the calculation. Remembering that the analysis of ref.\cite{FG1} leads to a vacuum energy density whose initial pulse seems destined to describe Inflation, one may ask if there is any simple and obvious way of removing those imaginary contributions to the unrenormalized $\Pi(k^2)$.

Again, the answer is positive, but comes at the price of an assumption which at first glance seems quite bold. If, in the quantum vacuum, in addition to every lepton and quark loop fluctuation, there also exists a corresponding massive, electrically charged fermionic--tachyon  loop fluctuation \cite{FG3}, it will contribute a negative imaginary term which exactly cancels that of the lepton or quark loop.

\bigskip
{\section{Dark Energy}}
\setcounter{equation}{0}

We first recall the essential features of the QED--based vacuum energy of ref.\cite{FG1}. Imagine a loop fluctuation, in which a virtual photon of the quantum vacuum splits into an electron and positron pair which exists for a short period of time before annihilating. While the charged $e^+$-- $e^-$ pair exists, there is ( thinking classically ) an electric field between them, and hence an electrostatic energy in that field, which fluctuation disappears when the $e^+$ and $e^-$ annihilate. But then another, and another, and more and more such fluctuations appear, with the normal to the plane of each loop in an arbitrary direction.

Traditionally, QED is formulated under the assumption that no electromagnetic fields can exist in the vacuum in the absencce of external charges and currents; and on macroscopic scales this certainly appears to be true. But on much smaller scales, and with correspondingly higher frequencies, there may be fluctuating fields containing energy, whose source is the fluctuating loops of the quantum vacuum, which -- on the average -- can serve to define a vacuum energy with interesting classical consequences. Ref.\cite{FG1} chooses to describe that average energy in terms of a C--number field ${A}_{\mu}^{\rm vac}(x)$, which resembles an external, classical field, except that it cannot be turned off; and a ``bootstrap'' equation is written for such an ${A}_{\mu}^{\rm vac}(x)$, in the sense that one first assumes it exists, and one then calculates its possible forms. It turns out that effective Lorentz invarianc is easily achieved -- each observer obtains the same form of solution to this vacuum energy in every Lorentz frame -- and the results are striking : the frequency of such an average fluctuation turns out to be on the order of $M_P\,c^2/\hbar$, the frequency associated with the Planck mass.

To derive a functional equation for ${A}_{\mu}^{\rm vac}(x)$, or rather for its Fourier transform $\tilde A_{\mu}^{\rm vac}(k)$, use is made of the second order, one--loop approximation to the vacuum functional $L[A]$ :
\beq L^{(2)}[A] \,=\, -{ i\over2}\int\!\!d^4x\,d^4y\,A^{\mu}(x)\,K^{(2)}_{\mu\nu}(x-y)\, A^{\nu}(y)\eeq
with $\tilde K_{\mu\nu}(k) = \bigl( g_{\mu\nu} \,k^2 - k_{\mu}k_{\nu}\bigr)\,\Pi(k^2)\ ,\ k^2 =  k_0^2 - \vec k^2$ and :
\beq\Pi(k^2) = {2\alpha_0\over\pi}\!\int_0^1\!\!dy\,y(1-y)\int_0^{\infty}\!\!{ds\over s}\,e\,^{\displaystyle -is[m^2-y(1-y)k^2]}\eeq 
where $m$ denotes the mass of the charged particle whose vacuum loops are the source of the vacuum field and $\alpha_0$ is the fine structure constant calculated with the bare, or unrenormalized charge $e_0$. 
The renormalized $\Pi_R(k^2)$ is given by :
\beq\Pi_R(k^2) = -{2\alpha\over\pi} \int_0^1\!\!dy\,y(1 - y)\ln\bigl[ 1 - y(1 - y){k^2\over m^2}\bigr]\eeq
where to this order, $\alpha$ is calculated with the renormalized charge $e_R$. Equations $(2.2)$ and $(2.3)$ have been obtained using Schwinger's manifestly gauge invariant formulation of QED \cite{HMF}; and the renormalized $(2.3)$ is exactly the same as that given by the more conventional approach, using Feynman graphs \cite{PS1}.

The equation which ref.\cite{FG1} obtains for $\tilde A^{\rm vac}$ takes the form :
\beq \Bigl(  \displaystyle{1 + 2\Pi_R(k^2)\over 1+\Pi_R(k^2)}\Bigr)\,\tilde A_{\mu}^{\rm vac}(k) = 0\eeq
and the most obvious solution is simply $\tilde A_{\mu}^{\rm vac}(k) = 0$, the conventional assumption. But there is another class of solutions, ``singular distribution solutions'', which take the form :
\beq \tilde A_{\mu}^{\rm vac}(k) =\,C_{\mu}(k) \,\delta(k^2 + M^2)\eeq
where both $C_{\mu}(k)$ and $M$ are to be determined. The mass term $M$ in  $(2.5)$ is the solution of the relation given by eq.$(2.4)$  :
\beq \Pi_R(-M^2) = -{1\over2}\eeq
As estimated in ref.\cite{FG1}, this purely QED calculation yields a value of $M\sim 10^{18}\,$GeV/c$^2$, the order of the Planck mass. As far as $C_{\mu}(k)$ is concerned, the choice of ref.\cite{FG1} is :
\beq C_{\mu}(k) =  v_{\mu}\,\delta(k\cdot v)\eeq
which enforces the Lorentz gauge condition $k\cdot\tilde A^{\rm vac}(k) = 0$, and where $v_{\mu}$ is a ``polarization'' vector. The fields and energy density associated to this solution are everywhere finite. Because the vacuum loops appear and re--appear, with the vectors normal to the plane of their loops taking all possible directions, the only sensible choice for such an averaged 4--vector must lie in the time direction; and this leads to an effective, electrostatic field ${\bf A}_{0}^{\rm vac}(x)$ :
\beq {\bf A}_{\mu}^{\rm vac}(x)\longrightarrow {\bf A}_0^{\rm vac}(r) \propto\,{\sin(Mr)\over r}\eeq
and an energy density $\rho =  \xi\,M^4f(x)$, with $x = Mr = Mct$, and $\xi$ a constant to be determined; here, $r$ and $t$ refer to space and time coordinates measured from the very origin of the Universe, and : 
\beq f(x) = {1\over x^2}\Bigl( \cos x - {\sin x\over x} \Bigr)^2\eeq

\begin{figure}
\includegraphics[width=10truecm]{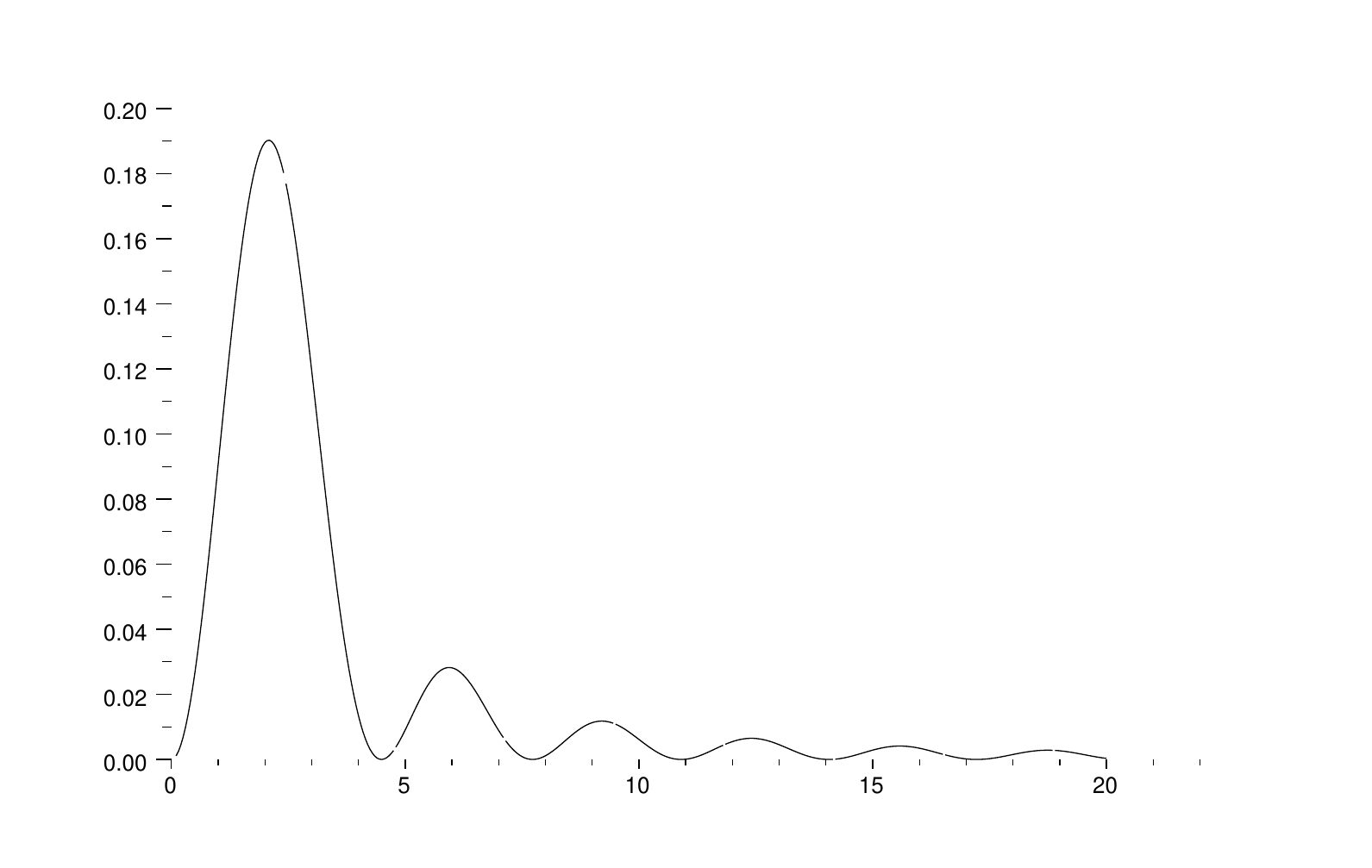}
\caption{A plot of $f(x) = \displaystyle{1\over x^2}\Bigl(\cos x - {\sin x\over x}\Bigr)^2$ vs. $x$}
\end{figure}

The plot of Fig.1 is interesting and suggestive. If we understand that we are living 13.7 billion years to the right of the origin of Fig.1, our present day, average $\rho_{\rm now}$ can be calculated by performing the volume integral over $\rho$ from $r=0$ to $r=R_{\rm now}$, where $R_{\rm now}$ denotes the present value of the radius of the Universe, and dividing by the volume of the present Universe. With the choice of parameter $\xi\sim1$, that average energy density $\rho_{\rm now}$ corresponds to a value on the order of $10^{-29}\,{\rm g/cm}^3$, which has been quoted \cite{LL}\cite{CP} as the value needed to produce the observed, outward acceleration of the Universe (the last measures from the Planck experiment \cite{CP} give $\rho = 0.578\,10^{-29}\,{\rm g/cm}^3$). Further details of this calculation may be found in ref.\cite{FG1}.

\bigskip
{\section{Inflation}}
\setcounter{equation}{0}
Intuitively, the first peak of Fig.1 suggests a large initial value of energy density which might well correspond to Inflation; and perhaps, all that is necessary would be to assign initial and final times, $t_i$ and $t_f$, halfway up and halfway down that first peak, and then calculate the corresponding $\rho$ at such initial and final times. But, as described in the Introduction, that procedure should be modified because electromagnetic charge renormalization cannot be defined without adequate spatial volume, which does not yet exist at the very beginning of the Universe. We of course admit that  QED is there -- eventually the only interaction at that time --  and that the photons can interact with leptons of mass $m_l$ and charge $\pm e_0$, even though we will consider the case of virtual lepton loops only. One must therefore employ the unrenormalized form of the loop function given in $(2.2)$, and for an electron--positron pair we have :
\beq\Pi(k^2) = {2\alpha_0\over\pi}\!\int_0^1\!\!dy\,y(1-y)\int_0^{\infty}\!\!{ds\over s}\,e\,^{\displaystyle -is[m_e^2-y(1-y)k^2]}\eeq

\indent As also noted in the Introduction, we shall replace ${2\alpha_0/\pi}$ by 1, following the analysis of ref.\cite{FG2}. The QED perturbative series coupling constant effectively being $\alpha_0/\pi$, it is still smaller than one, and a first order calculation in $\alpha_0/\pi$ is meaningful and can even be the only process available in such a small space--time situation. The logarithmic divergence of $(3.1)$ arising near $s=0$ will require a physical cut--off $\varepsilon$; and since the dimensions of $s$ are mass$^{-2}$, that lower cut--off  should signify that space--time distances smaller than the inverse Planck mass cannot be described within a physical theory that respects both Quantum Mechanics and Relativity : $\varepsilon = M_P^{-2}$, with $M_P = \displaystyle\sqrt{\hbar c\over G} \simeq 10^{19}$\,GeV$/c^2$.

The $s$ integral  is easy to perform, and we let  $\varepsilon = 0$ when no divergence appears :

\be \lim_{\varepsilon\rightarrow 0^+}\int_{\varepsilon}^{\infty}\!\!{ds\over s}\,e\,^{\displaystyle -is|A|} = \lim_{\varepsilon\rightarrow 0^+}\int_{\varepsilon}^{\infty}\!\!{ds\over s}\,\cos (s|A|)  -i\int_0^{\infty}\!\!{ds\over s}\,\sin (s|A|)  \ee
the values of the two integrals being well known :
\be \lim_{\varepsilon\rightarrow 0^+}\int_{\varepsilon}^{\infty}\!\!{ds\over s}\,\cos (s|A|) = - \ln(\varepsilon |A|)-\gamma \ \ \ \ \ \ {\rm and\,:}\ \ \ \ \int_0^{\infty}\!\!{ds\over s}\,\sin (s|A|)  = {\pi\over 2}\ee
where  $\gamma$ is Euler's constant $\simeq .577$. 

Coming back to eq.(3.1) and replacing $\varepsilon$ by $M_P^{-2}$, one obtains :
\be
\Pi(-M^2) =  - \int^1_0 \!\!dy\, y(1-y) \ln\biggl[{m_e^2\over  M^2_P}  +y(1-y){M^2\over  M^2_P}\biggr] - {\gamma\over6} -\frac{i\pi}{12}
\ee

\noindent in which the first and second RHS terms are clearly real, but the third term is
imaginary, conflicting with our model's demand that $\Pi(-M^2)$ be real. 

This unwanted imaginary term is, of course, not specific to the Schwinger's proper--time method employed here to calculate $\Pi(k^2)$ (eq.(3.1)). The unrenormalised loop computed from Feynman graphs (eq.(A.1)) using the gauge--invariant Dimensional Regularisation procedure also contains a finite imaginary term (A.4), as noted in the Appendix.

The inclusion of other lepton or quark loops ( if quarks do exist at that epoch ) on the RHS of (3.4)
 will only acerbate this situation, and we are lead to the question :
is there any way of eliminating such unwanted, imaginary
contributions, one that does not violate any principle of Quantum Field Theory~?

The answer is positive, and it is suggested by the
``tachyonic" argument of the delta function of $(2.5)$. Suppose that the
quantum vacuum, in addition to containing virtual, charged lepton--antilepton
loops also contains electrically charged, tachyon--antitachyon
loops, fermionic quantities which couple to photons in the same way as
do leptons, but whose internal symmetries are somewhat different \cite{FG3}.
Suppose that the Universe contains this built--in symmetry, so that for
every virtual fermion loop fluctuating in the quantum vacuum
there is a corresponding charged tachyon pair also fluctuating in that
vacuum. 

If such massive, charged tachyons existed, their 
dynamics could be described in a manner quite similar to fermions ( see ref.\cite{FG3} again ). The relevant statement here is that the
corresponding, closed tachyon loop functional, $L_T[A]$ takes on the
same form as that of the lepton or quark L[A], except for the change of sign of
its (mass)$^2$ term, where the $m^2$ of L[A] is changed to $-m_T^2$
inside $L_T [A]$.

The tachyonic contribution of each such virtual pair, to be added
to the lepton contribution of (3.4), is therefore : 

\be
\Pi_T(-M^2) =  
 - \int^1_0 \!\!dy\, y(1-y) \ln\biggl[{m_T^2\over  M^2_P}  - y(1-y){M^2\over  M^2_P}\biggr] - {\gamma\over6} +\frac{i\pi}{12} 
\ee

\noindent with the condition that the bracket of the logarithm be positive, which means $m_T > M/2$, for the maximum
value of $y(1-y)$ in the range $0<y < 1$ is 1/4. For $m_T < M/2$, we recover a negative imaginary part, like in (3.4). We also expect that $M$ will
be much larger than any of the fermions masses, which then
disappear from the problem, while it will be convenient to set $m_T = \eta M$,
where $\eta > 1/2$. Adding (3.4) and (3.5), we obtain :

\beq\matrix{\displaystyle\Pi_{l+T} (-M^2) &\displaystyle =  \,- \int^1_0 \!\!dy \,y(1-y) \bigg\{ \ln \bigg[\frac{m^2_\ell}{M^2_P}  + y(1-y) \frac{M^2}{M^2_P}\bigg]  
+ \ln \bigg[\frac{M^2}{M^2_P}  \Bigl(\eta^2 - y(1-y)\Bigr) \bigg] \bigg\}- {\gamma\over3}\hfill\cr\noalign{\medskip} & \displaystyle\simeq  {1\over3}\ln \big(\frac{M^2_p}{\eta M^2} \big)  + {5\over18} -{\gamma\over3} \,\hfill\cr\noalign{\medskip} & \displaystyle\simeq {1\over3}\Bigl[\,\ln \big(\frac{M^2_p}{\eta M^2} \big) + {1\over4} \,\Bigr]\hfill}\eeq

\noindent  If we add the remaining two lepton + tachyon loop
contributions, using the same, or an averaged value of $\eta$, we obtain :

\be
\Pi_{\sum{(l+T)}} (-M^2) \simeq \ln \big(\frac{M^2_p}{\eta M^2} \big) + {1\over4}
\ee
We can also compute the contributions of the quarks -- if they are present -- and their associated tachyons. For a quark of electric charge $e_q|e_0|$, with $e_q = 2/3$ or $e_q = -1/3$, this quark + tachyon loops will give :
\be
\Pi_{q+T} (-M^2) \simeq 3\times{e^2_q\over3}\Bigl[\,\ln \big(\frac{M^2_p}{\eta M^2} \big) + {1\over4} \,\Bigr]
\ee
where the color degree of freedom has been taken into account and we use the same $\eta$ as previously. The total contribution of the three charged leptons + six quarks + tachyons would then give :
\be
\Pi_{\sum{(l+q+T)}} (-M^2) \simeq {8\over3}\Bigl[\,\ln \big(\frac{M^2_p}{\eta M^2} \big) + {1\over4} \,\Bigr]
\ee

This tachyonic assumption cannot be ruled out by any existing experiment or observation; and it leads to
a coherent picture of Inflation and Dark Matter, in addition to Dark Energy, as
well as to a possible understanding
 of the Why and the How of the Big Bang, as noted in Section 5.
\bigskip
{\section{Computation}}
\setcounter{equation}{0}

When we set $\Pi (-M^2) $ equal to $-1/2$, as required by this QED--based
 model of Vacuum Energy, we obtain, using (3.7) ( just leptons )  :

\be
\frac{M_p}{\sqrt\eta M} \simeq 0.7
\ee
and we obtain, using (3.9) ( leptons and quarks ) :
\be
\frac{M_p}{\sqrt\eta M} \simeq 0.8
\ee
\noindent Clearly, the number of fermion + tachyon pairs doesn't really matter, and the nature of the fermions doesn't really matter either -- quarks can be absent from this scenario --, and we shall take : $M\simeq M_P/\sqrt{\eta}$,

If we now associate the leading peak of Fig.1 to the Inflation process, we see that it begins near $x = 1$, will be in full swing during the growing half of that
first pulse, and decreases during the second half of that pulse, ending
near $x = 4$.

We shall simply choose the Liddle and Lyth \cite{LL} parameters
for the initial time of inflation $t_i$ , and the value (in units of GeV) for the
initial energy density $\rho_i$, parameters stated without any uncertainty;
and then choose our two parameters to match the end time of
Inflation $t_f$, associated with the average energy density at that time, $\rho_f$.
If the initial parameters are allowed any reasonable variations, there
could then be corresponding variations of the model parameters.

Following reference \cite{LL}, we therefore choose :

\be
t_i = \frac{1}{c} \frac{1}{M_P} \rightarrow \frac{1}{c} \bigg(\frac{\hbar}{M_P c} \bigg) = \sqrt{\hbar G\over c^5} \simeq 10^{-43}\,{\rm s}
\ee

\noindent and from the model write the expression for $t_f$ in terms of $\eta$ :

\be
t_f = \frac{1}{c} R_f = \frac{x_f}{cM} \simeq \frac{4}{c} \bigg(\frac{\hbar}{Mc} \bigg) \simeq \frac{4}{c} \bigg(\frac{\hbar}{M_pc}  \bigg) \sqrt{\eta} \simeq 10^{-43} \sqrt{\eta}\,{\rm s}
\ee

\noindent Setting this equal to the Liddle and Lyth value of $10^{-32\pm 6}$ yields $\sqrt{\eta} = 10^{11\pm 6}$. We then obtain $M\sim 10^{8\pm 6} $\,GeV$/c^2$, and $m_T \sim 10^{30\pm 6}$\,GeV$/c^2$.

Inspection of Fig. 1
shows that the energy density in that first pulse is roughly $0.1\,\xi \, M^4$, or $\rho^{1/4}_f \sim 10^{-1/4}\,\xi^{1/4} M \sim \xi^{1/4}\,10^{8\pm 6} GeV$, which is compatible with the bounds of the listed $10^{13 \pm 3}$ if one chooses $\xi\sim O(1)$. In this way, with the same parameter as needed for the Dark Energy estimation, this QED
vacuum energy Model can satisfy the cosmological requirements for both Inflation and Dark Energy.

And one may wonder if there is a cosmological significance to
a tachyon mass far exceeding $M_P$. There are several implications of such a result. One of the consequences of that enormous mass, is that tachyons can acquire an extreme speed, and so travel a ``long'' distance during the duration of Inflation. We can evaluate this length using the assumption that the tachyon energy is given by the ``vacuum loop'' tachyonic mass $M$ : $E_T = Mc^2$. Recalling the relation between energy $E_T$ and speed $v_T$ for a tachyon \cite{FG3} : 
$$E_T = \displaystyle{m_Tc^2\over\sqrt{\displaystyle{v_T^2\over c^2} - 1 }}$$
we obtain an average speed $v_T = 10^{30}$\,ms$^{-1}$. If we take $t = t_f =10^{-32}$\,s as the travel time, then the distance covered by the tachyon/antitachyon will be $R = 10^{-2}$\,m, which is an admitted value for the radius of the Universe at the end of Inflation \cite{PS}. So to speak, we could say that tachyons ``open up'' space--time, in all directions, and drag along photons with them.
\bigskip
{\section{A Cosmological Speculation}}
\setcounter{equation}{0}

Let us continue to explore the consequences of a huge tachyon mass
. We
shall still adhere to the concept that the smallest coordinate differences
compatible with GR and QM are given by the
inverse of the Planck Mass. But what we shall imagine here is that the
initial vacuum energy deposited as the new Universe appears is not
the Planck mass, with an initial energy density of $M_P^4$; but rather
that this happy event was generated by the annihilation of a random $T-\bar{T}$
pair with total energy of a few $ m_T$ (to be conservative).

Then the $t_{i,f}$and $R_{i,f}$ calculations go through as before, because
those estimates were based on the inverse Planck length, to which we
continue to adhere as the smallest possible coordinate difference. But
at the end of Inflation, one would now find :
$$\rho_f \sim  {m_T\over{\displaystyle{4\pi\over3} R^3_f }}$$
Using $\displaystyle R_f\sim {x_f\over M}$, one obtains :
$$\rho_f \sim {m_TM^3\over x^3} \sim 10^{52}\,{\rm GeV}^4\ \ \ \longrightarrow\ \ \ \rho_f^{1/4} \sim 10^{13}\,{\rm GeV}$$
 which is the value quoted in Liddle and Lyth. To the authors, this seems like a rather more elegant way of fitting data,
and it may perhaps suggest a somewhat daring Cosmological Speculation, as follows.

Suppose that a sizable amount of energy, much larger than $M_p$,
is suddenly deposited -- for example, by the accidental annihilation of a
highly energetic $T -\bar{T}$ pair -- at one point in a coordinate system whose
space-time structure cannot support that much energy. What may well
happen at that point is that a new coordinate system, of a New
Universe, appears with no memory of its origin, and a corresponding
Inflation begins. At that point, or in the extremely small region in which
this occurs, one may imagine that the Old Universe's space-time
structure has been ``torn", or disrupted in such a way that -- in that
region only -- the separation of vacuum energy from real energy is
disrupted, and that the immense amount of the Old Universe's vacuum
energy is able to force its defining vacuum fluctuations through that
tear, and in the process convert them to real lepton--anti lepton, real
quark--anti quark, and real tachyon--anti tachyon pairs. In brief, this is
the Big Bang of the New Universe, with the electron--positron pairs
appearing first, because they are the lightest.

One can also think of this as a spectacular Schwinger Mechanism,
wherein the potential energy of the quantum vacuum is able -- at that
point -- to convert to ``real" energy, and then tear the vacuum
fluctuations, which originally defined that energy, out of the Old
Universe's vacuum energy. The details, of course, can only be
imagined; even for terrestrial events, it is difficult to describe an
explosion in terms of probabilistic effects. But one may note that this
vision of a New Universe's Inflation and its Big Bang will satisfy
conservation of energy and electrical charge.

Finally, it is difficult to refrain from considering the fate of the Old
Universe, if that entity continues to lose a sizable portion of its vacuum
energy to the New Universe. For it is, in this Model of Inflation and
Dark Energy, the potential energy locked in the quantum vacuum
which serves to resist the mutual gravitational attraction of a universe's
many parts; and if this vacuum energy, or a too large portion of it, is
lost to the New Universe, the result must be the collapse of the Old
Universe into a monstrous black hole, one whose radiation could well
be observable astronomically, by Astrophysicists of the New Universe.

\bigskip
{\section{Summary}}
\setcounter{equation}{0}

The above paragraphs come at the
end of a line of intuitive reasoning, which began with a new prediction for a possible QED vacuum energy. Tachyons were
introduced when demanded by the model, in order to remove an
unwanted, imaginary contribution. To conclude, this model, with one arbitrary parameter, is able to describe both Dark Energy and Inflation.

It may also be useful to note that this model of Inflation is expressed in terms of physical fluctuations, which are the way in which many physical changes take place; and that the spaces between successive pulses -- especially the first and the second -- may have physical significance. One can perhaps answer the question as to why an observed rate of universe expansion falls to a value lower than that presently observed; and the answer could well be that the first pulse of Fig.1 defines the initial burst of Inflation, but that during the trough between the first and the second pulse, the inflation falls to a rate lower than or on the order of the present day rate. There could well be other effects of each successive pulse, starting for example with the onset of the electroweak epoch.

\vskip1truecm 
\appendix
\def\theequation{{A}.\arabic{equation}}
{\bf\Large Appendix}

\vskip0.3truecm 
We want to show in this Appendix how the usual approach to the photon self--energy computation, at one loop level, using Feynman graphs techniques, leads to the same feature as the one obtained with the Schwinger proper--time method.

The first order loop $\Pi^{\mu\nu}(k)$ is given by :
\be i\Pi^{\mu\nu}(k) = (-ie)^2\int\!{d^4p\over(2\pi)^4}\,{\rm Tr}\left(\gamma^{\mu}{i\over \slash \!\!\!p - m}\gamma^{\nu}{i\over \slash \!\!\!p - \slash \!\!\!k - m}\right)\ee
Using dimensional regularisation (DR), we find \cite{PS1} :
\be \Pi^{\mu\nu}_{DR}(k) = {2\alpha\over\pi}\bigl( g^{\mu\nu} \,k^2 - k^{\mu}k^{\nu}\bigr)\left\{{1\over3\varepsilon} - \int_0^1\!dy\,y(1 - y)\ln\biggr[{-m^2 + y(1 - y)k^2 \over\mu^2}\biggr] - {\gamma\over6} + O(\varepsilon)\,\right\}\ee
where $\varepsilon = 4 - d$ and $\varepsilon\rightarrow0^+$ ($1/\varepsilon$ takes care of the logarithmic divergence of $\Pi$), and $\mu$ is an arbitrary mass parameter introduced to get the correct dimension of the lagrangian when $d\ne4$.

Introducing the Planck mass by setting $\displaystyle{1\over\varepsilon} = \ln{M_P\over\mu}$, one immediately gets :
\be \Pi_{DR}(-M^2) =  - \int_0^1\!dy\,y(1 - y)\ln\biggr[{-m^2 - y(1 - y)M^2 \over M_P^2}\biggr] - {\gamma\over6} \ee
Taking out the minus sign inside the log leads to an imaginary term : 
\be \Pi_{DR}(-M^2) =  - \int_0^1\!dy\,y(1 - y)\ln\biggr[{m^2 + y(1 - y)M^2 \over M_P^2}\biggr] - {\gamma\over6} - {i\pi\over6}\ee

\vskip1truecm {\bf\Large Acknowledgement}
\vskip0.3truecm 
It is a pleasure to acknowledge helpful conversations with W. Becker, S. Koushiappas and P.D. Mannheim.

This publication was made possible through the support of the
Julian Schwinger Foundation.

\end{document}